\documentclass[%
 reprint,
 amsmath,amssymb,
 aps,
]{revtex4-2}

\usepackage{graphicx}
\usepackage{dcolumn}
\usepackage{bm}
\usepackage{xcolor}
\usepackage{subcaption}
\DeclareMathOperator\arctanh{arctanh}

\begin{document}

\preprint{APS/123-QED}

\title{Dissipation in spin chains using quantized nonequilibrium thermodynamics }

\author{Massimo Borrelli}
\email{massimo.borrelli@mat.ethz.ch}
\author{Hans Christian \"{O}ttinger}

\affiliation{Polymer Physics, Department of Materials, ETH Z\"{u}rich, CH-8093 Z\"{u}rich, Switzerland}%




\date{\today}

\begin{abstract}
We investigate the open dynamics of a chain of interacting spins using the quantized version of the GENERIC equation from classical out-of-equilibrium thermodynamics. We focus on both equilibrium and nonequilibrium scenarios for chains of different sizes. While in the equilibrium case we demonstrate thermal equilibration to the correct many-body Gibbs density matrix, in the nonequilibrium dynamics we show a ballistic-to-diffusive transition in the steady-state energy current and a scaling that is consistent with Fourier's law of heat transfer. 
\end{abstract}

\maketitle

\section{Introduction}  Until a few years back the study of open quantum dynamics was mostly concerned with systems in quantum optics and atomic physics \cite{breuer_petr,weiss}. Nowadays, thanks to major technical improvements in experimental control, a great deal of attention is devoted to simulating dissipation in engineered condensed matter systems \cite{Capriotti_2002,Di-Candia:2015wv,Weimer:2010vq,sim_dyn_map_ions}. In this, respect, lattice models such as Heisenberg and Hubbard \cite{heisenberg,hubbard_rev} are paramount examples of many-body dissipation \cite{PhysRevA.84.031402,PhysRevLett.110.233601}. 

The general framework within which these investigations are typically conducted is that of Markovian master equations in the standard Lindblad form and it has been applied to both driven and autonomous systems \cite{foss2017solvable,PhysRevB.98.241108,PhysRevA.98.052109}. For a recent review, see \cite{sim_open_systems_rev}. This approach has certainly several advantages. The Lindblad master equation is linear and can be solved using well-established techniques, such as quantum jumps \cite{Weimer_2016} and Quantum Monte Carlo methods \cite{nagy_2018}. Moreover, numerical methods originally designed for isolated systems, {\itshape e.g.} tensor network methods \cite{Orus:2019vt} and linked cluster expansions \cite{TANG2013557}, have been sucessfully extended to the open quantum system domain \cite{PhysRevLett.93.207204,PhysRevLett.93.207205,PhysRevB.97.035103}. Another important aspect of the Lindblad master equation approach lies in the spectral properties of the Liouvillian operator. By calculating its eigenvalues and eigenoperators one can investigate the steady-state properties of the many-body system, such as Green's functions, relaxation rates \cite{PhysRevE.92.042143,Prosen_2010}, and even critical features \cite{PhysRevA.98.042118,PhysRevB.98.241108}. 

While the Markovian approximation can be considered completely satisfactory whenever no memory effects are present, if the system-environment coupling is not weak and the Bohr frequencies of the system are not well separated (thus leading to a breakdown of the secular approximation), the use of a local Lindblad master equation can lead to some difficulties \cite{Levy_2014,Trushechkin_2016,local_global,Mitchison_2018,Cattaneo_2019}. First of all, in order for the Lindblad equation to be thermodynamically consistent in predicting steady state properties, the Hamiltonian spectrum and its eigenstates should be fully known which is rarely the case in many-body physics, where Hamiltonians are in general very complex and can be expected to have a dense energy spectrum. Thus, very often, when studying the open dynamics of many-body systems, local Lindblad operators are assumed, describing quantum jumps between eigenstates of the non-interacting part of the Hamiltonian. This single-particle approach works reasonably well for weakly interacting many-body systems. However, as most systems in condensed matter physics are highly correlated, a single-particle picture of dissipation could be misleading and a collective formulation of the Lindblad operators should instead be considered in order to predict equilibration and steady-state properties correctly. This route has been and is still being explored in the literature, often in conjunction with attempts to relax the secular approximation and has led to the formulation of many-body Redfield equations describing thermodynamic properties correctly \cite{PhysRevE.61.2397,Saito_2003,Davidovic2020completelypositive}. However, when using Redfield equation, the main drawback is the possibility of generating negative quantum probabilities, even though this is mostly true in the short-time regime.

Here, we explore a completely different approach to the study of dissipative many-body physics. Instead of microscopically deriving or explicitly constructing a Markovian master equation in Lindblad or Redfield form, we model dissipation via a thermodynamic nonlinear master equation that was first presented in \cite{Grabert:1982ty} and, later on, mathematically formalized and generalized using a geometric approach in \cite{_ttinger_2011,nat_diss}. This equation was originally proposed as the quantum generalisation of the GENERIC (general equation for the nonequilibrium reversible-irreversible coupling) framework of classical out-of-equilibrium thermodynamics \cite{PhysRevE.56.6620,PhysRevE.56.6633,beyond_eq}. Conceptually, it relies on a clear-cut separation of reversible versus irreversible dynamics, the former being generated by the Hamiltonian, the latter being driven by the system entropy. This separation translates to different geometric structures in the equation. While the reversible part is associated to the standard commutator of the density matrix with the system's Hamiltonian, the irreversible part leads to a term that is nonlinear in the density matrix but guarantees thermodynamic consistency and correct steady-state properties. This master equation was successfully applied to traditional open system scenarios, such as spontaneous decay of a two-level atom and the Caldeira-Legget model where numerical solutions were found by using deterministic integration methods for nonlinear equations as well as adaptations of stochastic unraveling \cite{PhysRevA.82.052119,PhysRevA.86.032101,PhysRevA.86.032102,Osmanov:2013ut,HansZeroTemp}. 

Here, for the first time, we apply this framework to the study of dissipation in a many-body system, that is, an open spin chain described by the XXZ model. Although the system-environment coupling operators are local, no jump operators are needed at any point. This lifts the intrinsic ambiguity in choosing the correct representation, {\itshape i.e.}, global versus single-particle, and no assumptions regarding the spectrum of the chain are required. We apply and solve GENERIC to equilibrium dynamics, where we demonstrate full equilibration to the correct many-body Gibbs ensemble, and to a nonequilibrium scenario as well, where we recover a transition from ballistic to diffusive energy transport, in agreement with previous studies on related models \cite{Michel:2003vv,PhysRevE.86.061118,PhysRevLett.106.217206,PhysRevB.80.035110,PhysRevB.87.235130,Mendoza_Arenas_2013}. 

This manuscript is organised as follows. In section \ref{sec:model} we introduce the general $N-$spin model and, within the GENERIC framework, derive a nonlinear master equation describing dissipative dynamics. In section \ref{sec:two_spins} we calculate the steady-state of our nonlinear master equation analytically for 2-spin chain toy-model and show that it is the correct Gibbs state. In section \ref{sec:solution} we solve the nonlinear master equation for several spin chain lengths, prove thermalisation and calculate thermalisation rates. In section \ref{sec:ooe} we consider an out-of-equilibrium scenario where a shorter chain is connected at its extremities to two distinct heat baths at different temperatures. Finally, in section \ref{sec:conclusions} we draw some conclusions and discuss open questions, perspectives and possible future directions.
 
\section{The open XXZ GENERIC model}
\label{sec:model}
We consider a chain of $N$ interacting $1/2$ spins in an open configuration whose isolated dynamics is dictated by the following Heisenberg-type Hamiltonian (assuming $\hbar=1$)
\begin{equation}
H = \sum_{j}\sum_{j = 1}^{N} \tau_{j}\sigma_{j}^{k}\sigma_{j+1}^{k},
\label{general_H}
\end{equation}
where $k=x,y,z$. Eq.~\eqref{general_H} can be simplified if one sets $\tau_{x}=\tau_{y}=\tau$, $\tau_{z}=\Delta$  leading to
\begin{equation}
H = \tau\sum_{j = 1}^{N} \left(\sigma_{j}^{+}\sigma_{j+1}^{-}+\sigma_{j}^{-}\sigma_{j+1}^{+}\right)+\Delta\sum_{j = 1}^{N} \sigma_{j}^{z}\sigma_{j+1}^{z},
\label{simpler_H}
\end{equation}
which, in literature, goes under the name of Heisenberg XXZ model. We rename the two contributions as $H^{\textrm{coll}}=\tau\sum_{j = 1}^{N} \left(\sigma_{j}^{+}\sigma_{j+1}^{-}+\sigma_{j}^{-}\sigma_{j+1}^{+}\right),$ and $H^{\textrm{free}}=\Delta\sum_{j = 1}^{N} \sigma_{j}^{z}\sigma_{j+1}^{z}.$ For now, we assume that all the spins are locally coupled to a heat bath at temperature $T$. 

To describe the open dissipative dynamics of the spin chain we use the following thermodynamic quantum master equation \cite{PhilApp} which we named GENERIC
\begin{equation}
\begin{aligned}
\dot{\rho_{t}} = i\left[\rho_{t},H\right]&-\sum_{j = 1}^{N}\int_{0}^{1}du f_{j}(u)[Q_{j},\rho_{t}^{1-u}[Q_{j}^{\dagger},\mu_{t}]\rho_{t}^{u}]\\
&-\sum_{j = 1}^{N}\int_{0}^{1}du f_{j}(u)[Q^{\dagger}_{j},\rho_{t}^{u}[Q_{j},\mu_{t}]\rho_{t}^{1-u}],
\label{generalME}
\end{aligned}
\end{equation}
where the operators $Q_{j}$ model the system-environment weak coupling and the $f_{j}(u)$ are real non-negative rate factors. Eq.~\eqref{generalME} was first introduced in \cite{_ttinger_2011} and it is based on a fundamental, postulated separation between reversible and irreversible dynamics. Reversibility is accounted for by the standard commutator between $\rho_{t}$ and $H$. The nonlinear term generates the irreversible part of the dynamics through the double commutator and the free energy operator $\mu_{t}=H+k_{B}T\log\rho_{t}$. Unlike standard master equations in quantum theory this equation is not derived microscopically, but rather postulated and constructed as the quantum generalisation of the classical nonequilibrium GENERIC equation. For more technical details, see \cite{_ttinger_2011,nat_diss}. Thus, irreversibility \textit{is not} derived from course-grained reversible quantum behaviour; it is included \textit{a priori} and associated to a precise geometrical structure (double commutator) and a precise generator (free energy). Another way to look at Eq.~\eqref{generalME} is to interpret it as a phenomenological equation. Although the $Q_{j}$ operators describe how the open system couples to its surrounding environment, no full knowledge of the system-environment interaction Hamiltonian is required. In other words, we only need to care about the system-environment coupling from the system's perspective. Obviously, one has to both chose sensible couplings $Q_{j}$ and model the rates $f_{j}(u)$ properly which is not at all an \textit{a priori} obvious task. Yet, because of its explicit geometric construction based on thermodynamic arguments, Eq.~\eqref{generalME} should always predict the correct steady-state behaviour, assuming that proper modelling of both rates and coupling is realized \cite{nat_diss}. We would also like to remark that Eq.~\eqref{generalME} describes Markovian dynamics only.

Here, we assume that the spins couple to the heat bath via their $z-$components $\sigma^{z}$. The reason for this choice simply lies in the fact that, as we will see later, it will allow us to restrict our attention to subspaces of the total Hilbert space with fixed total spin $z$-component, that is $S_{z}=\sum_{j=1}^{N}\sigma_{j}^{z}$. 
If we were to draw a loose parallel with a traditional approach starting from a system-environment interaction Hamiltonian, we might guess this would be of a spin-boson type, that is, $H_{j}=g_{j}\sigma_{\alpha_{z}}\sum_{k}(a_{k}+a^{\dagger}_{k})$ with the environment being in a thermal state $\rho_{E}\propto\Pi_{k}e^{-\beta\omega_{k}a^{\dagger}_{k}a_{k}}$. However, as we would like to stress that this is purely a speculation, only rooted in the type of spin-bath coupling. Hence, from now on $Q_{j}=\sigma_{j}^{z}=Q_{j}^{\dagger}$ and equation \eqref{generalME} simplifies greatly
$$
\begin{aligned}
\dot{\rho_{t}} = i\left[\rho_{t},H\right]&-2\sum_{j = 1}^{N}\int_{0}^{1}du f_{j}(u)[\sigma_{j}^{z},\rho_{t}^{1-u}[\sigma_{j}^{z},\mu_{t}]\rho_{t}^{u}]
\end{aligned}
$$
where we have used the invariance of the integrand under the transformation $u\to 1-\lambda$, which is guaranteed provided that $f_{j}(u)=f_{j}(1-u)$. We choose $f_{j}(u)=f_{j}$ constant, thus fulfilling such condition. In \cite{HansZeroTemp} it was shown that for a single two-level system described by $H^{\textrm{free}}=\omega\sigma_{z}$, the rate $f(u)$ can be chosen as $\exp(2u\omega/k_{B}T)$ if $Q=\sigma^{+}$. The same reasoning was later extended to a multi-mode bosonic field with $Q_{k}=a_{k}$ \cite{PhilApp}. In our model, the system-environment coupling is realised via the $\sigma^{z}$ operator and since $[\sigma^{z}_{j},H^{\textrm{free}}]=0$, the $f_{j}(u)$ functions can be constants. The dissipators can be further simplified if one notices the following
$$
\begin{aligned}
&\rho_{t}^{1-u}[\sigma_{j}^{z},\mu_{t}]\rho_{t}^{u} = \\
&\rho_{t}^{1-u}[\sigma_{j}^{z},H^{\textrm{free}}+k_{B}T\log\rho_{t}]\rho_{t}^{u}+\rho_{t}^{1-u}[\sigma_{j}^{z},H^{\textrm{coll}}]\rho_{t}^{u}=\\
&\frac{1}{\beta}\frac{d}{du}\left(\rho^{1-u}_{t}\sigma_{j}^{z}\rho^{u}_{t}\right)+\rho_{t}^{1-u}[\sigma_{j}^{z},H^{\textrm{coll}}]\rho_{t}^{u}
\end{aligned}
$$
with $\beta = 1/k_{B}T$ which, once integrated, leads to the following equation
$$
\begin{aligned}
\dot{\rho_{t}} =  i\left[\rho_{t},H\right]&-\sum_{j = 1}^{N}\frac{f_{j}}{\beta}\left(\rho_{t}-\sigma_{j}^{z}\rho_{t}\sigma_{j}^{z}\right)\\
&-2\sum_{j = 1}^{N}f_{j}\int_{0}^{1}du [\sigma_{j}^{z},\rho_{t}^{1-u}[\sigma_{j}^{z},H^{\textrm{coll}}]\rho_{t}^{u}]
\end{aligned}
$$
The linear part is formally identical to a standard dephasing bosonic bath with an Ohmic spectrum whose action can be described by a local Lindblad master equation, with a dephasing rate proportional to the temperature of the bath $\beta^{-1}$. 
On the second line, however, we find a nonlinear contribution that can be calculated and simplified to some extent. The first nested commutator can be calculated easily  
$$
\begin{aligned}
[\sigma_{j}^{z},H^{\textrm{coll}}] &=2\tau\left(-\sigma_{\alpha-1}^{+}\sigma_{j}^{-}+\sigma_{\alpha-1}^{-}\sigma_{j}^{+}+\sigma_{j}^{+}\sigma_{j+1}^{-}-\sigma_{j}^{-}\sigma_{j+1}^{+}\right) \\
&=2\tau\left(-S_{\alpha-1}+S_{j}\right),
\end{aligned}
$$
where we introduced the operators $S_{j}=\sigma_{j}^{+}\sigma_{j+1}^{-}-\sigma_{j}^{-}\sigma_{j+1}^{+}$. Note that these operators conserve the total spin along the $z$-direction, {\itshape i.e.} $S_{z}=\sum_{j=1}^{N}\sigma_{j}^{z}$. After some rearrangements, we arrive at the following equation
\begin{equation}
\begin{aligned}
\dot{\rho_{t}} =  i\left[\rho_{t},H\right]&-\frac{4}{\beta}\sum_{j = 1}^{N}f_{j}\left(\rho_{t}-\sigma_{j}^{z}\rho_{t}\sigma_{j}^{z}\right)\\
&-4\tau\sum_{j = 1}^{N}[\Delta\sigma_{j}^{z},S_{j}^{\rho_{t}}],
\end{aligned}
\label{FinalGenericEq}
\end{equation}
where we have introduced the shorthand notation $A^{\rho}=\int_{0}^{1}du\rho^{1-u}A\rho^{u}$ and the operators $\Delta\sigma_{j}^{z}=f_{j}\sigma_{j}^{z}-f_{j+1}\sigma_{j+1}^{z}$. 

Without any further simplifying assumption this is as far as we can get in deriving a GENERIC-type master equation describing the spin-chain with a local $\sigma_z$-type dissipative coupling to a heat bath. Obviously, the nonlinear term represents a difficulty in solving this equation as one has to diagonalise $\rho_{t}$ at every time-step. However, as we will show later, for some initial conditions this can be done numerically for fairly large chain sizes. If $p_{n}$ and $|\pi_{n}\rangle$ are the eigenvalues and eigenvectors of $\rho_{t}$ respectively then the commutators in the nonlinear part of master equation \eqref{FinalGenericEq} read
\begin{equation}
\begin{aligned}
[\Delta\sigma_{j}^{z},S_{j}^{\rho_{t}}]=\sum_{n,m}&\left(\frac{p_{n}-p_{m}}{\log p_{n}-\log p_{m}}\right)\times\\
&\langle\pi_{n}|S_{j}|\pi_{m}\rangle[\Delta\sigma_{j}^{z},|\pi_{n}\rangle\langle\pi_{m}|].
\label{nonLinear}
\end{aligned}
\end{equation}
As anticipated earlier, all the terms in Eq.~\eqref{FinalGenericEq} conserve the total spin along the $z-$direction. Thus, the bath induces incoherent transitions between eingenstates of $S_{z}$. That means that if the initial state $\rho_{0}$ is an eigenstate of $S_{z}$, so will be $\rho_{t}$. This, in turn, will translate to a great reduction of computational resources needed to solve Eq.~\eqref{FinalGenericEq}; by initialising the state of the chain to an eigenstate of $S_{z}$, we will reduce the complexity of the problem.

\section{A toy model: the $N=2$ case}
\label{sec:two_spins} 
In this section we will look at the simplest case of $N=2$ interacting spins. We will analytically calculate the steady state $\rho_{ss}$ of Eq.~\eqref{FinalGenericEq} in the $S_{z}=0$ subspace and show that it is precisely the Gibbs state with respect to Hamiltonian \eqref{simpler_H}, that is
\begin{equation}
\rho_{ss}=\rho_{G}=\frac{1}{\mathcal{Z}}e^{-\beta H}.
\label{gibbs}
\end{equation}
The two remaining states $|\uparrow,\uparrow\rangle$ and $|\downarrow,\downarrow\rangle$ are eigenstates of Hamiltonian~\eqref{general_H} and, being that this can be block-diagonalized, the inclusion of these states in the Gibbs state calculation is trivial. 
For two spins the XXZ Hamiltonian reduces to
\begin{equation}
H = \tau\left(\sigma_{1}^{+}\sigma_{2}^{-}+\sigma_{1}^{-}\sigma_{2}^{+}\right)+ \Delta\sigma_{1}^{z}\sigma_{2}^{z}
\label{H_2_spins}
\end{equation}
For the sake of simplicity in the calculations that follow, we set $\Delta=\tau$. The physics described by the above system is quite simple; the spin-spin interaction couples the states $|\uparrow,\downarrow\rangle$ and $|\downarrow,\uparrow\rangle$ to each other whilst leaving $|\uparrow,\uparrow\rangle$ and $|\downarrow,\downarrow\rangle$ unaffected. The eigenvalues are $E_{n}=0, -2 \tau, \tau, \tau$, the first two corresponding to eigenstates with $S_{z}=0$ that are linear combinations of $|\uparrow,\downarrow\rangle$ and $|\downarrow,\uparrow\rangle$.
Eq.~\eqref{FinalGenericEq} in this case reads
\begin{equation}
\begin{aligned}
\dot{\rho_{t}}=i[\rho_{t},H_{2}]&-\frac{4}{\beta}\left[(f_{1}+f_{2})\rho_{t}-f_{1}\sigma_{1}^{z}\rho_{t}\sigma_{1}^{z}-f_{2}\sigma_{2}^{z}\rho_{t}\sigma_{2}^{z}\right]\\
&-4\tau[\Delta\sigma_{z},S^{\rho_{t}}],
\label{generic2}
\end{aligned}
\end{equation}
where $\Delta\sigma_{z}=f_{1}\sigma_{1}^{z}-f_{2}\sigma_{2}^{z}$. As the $S_{z}=0$ subspace is two-dimensional, we can use an effective two-level description since we know that the total dissipative dynamics will not mix different subspaces. Therefore, we assume the following general form for $\rho_{0}$
\begin{equation}
\rho_{0}=\left(
\begin{array}{cc}
p  &  c  \\
c^{*}  & 1-p  
\end{array}
\right).
\label{rho0}
\end{equation}
The time-evolved state $\rho(t)$ will be always of the same form as $\rho_{0}$. Eq.~\eqref{generic2} can recast in a more concrete form that emphasises its nonlinear nature. For a general $\rho$ such as the one in Eq.~\eqref{rho0} the eigenvalues $p_{\pm}$ and eigenvectors $|\pi_{\pm}\rangle$ can be calculated analytically
\begin{equation}
p_{\pm}=\frac{1\pm p_{0}}{2},\;|\pi_{\pm}\rangle=\frac{1}{\sqrt{N_{\pm}}}\left[|\uparrow,\downarrow\rangle+\left(\frac{p_{\pm}-p}{c}\right)|\downarrow,\uparrow\rangle\right],
\label{rho_eigen}
\end{equation}
where  $p_{0}=\sqrt{1+4\left[|c|^{2}+p(p-1)\right]}$ and $N_{\pm}$ is a normalisation factor. The commutator $[\Delta\sigma_{z},S_{\rho}]$ can be analytically calculated using Eq.~\eqref{nonLinear} 
$$
\begin{aligned}
[\Delta\sigma_{z},S^{\rho_{t}}]&=\left(\frac{p_{+}-p_{-}}{\ln p_{+}-\ln p_{-}}\right)\langle S\rangle_{+-}[\Delta\sigma_{z},\Pi_{+-}]\\
&+\left(\frac{p_{-}-p_{+}}{\ln p_{-}-\ln p_{+}}\right)\langle S\rangle_{-+}[\Delta\sigma_{z},\Pi_{-+}]\\
&+p_{+}\langle S\rangle_{++}[\Delta\sigma_{z},\Pi_{++}]+p_{-}\langle S\rangle_{--}[\Delta\sigma_{z},\Pi_{--}],
\end{aligned}
$$
where $\langle S\rangle_{\alpha\beta}=\langle\pi_{j}|S|\pi_{\beta}\rangle$ and $\Pi_{\alpha\beta}=|\pi_{j}\rangle\langle\pi_{\beta}|$. All the terms in the above equation can be evaluated explicitly. After some extra steps one arrives at the following equation
\begin{equation}
\begin{aligned}
\dot{\rho_{t}}&=i[\rho_{t},H_{2}]-\frac{4}{\beta}\left[(f_{1}+f{_2})\rho_{t}-f_{1}\sigma_{1}^{z}\rho_{t}\sigma_{1}^{z}-f_{2}\sigma_{2}^{z}\rho_{t}\sigma_{2}^{z}\right]\\
&-4\tau(f1+f2)\left[\lambda(p,c)|\downarrow,\uparrow\rangle\langle\uparrow,\downarrow|+\lambda^{*}(p,c)|\uparrow,\downarrow\rangle\langle\downarrow,\uparrow|\right],
\label{generic2level}
\end{aligned}
\end{equation}
where $\lambda(p,c)=\lambda(p(t),c(t))$ is a function of the state $\rho_{t}$ and fully determines the structure of the nonlinear part of the irreversible dynamics
\begin{equation}
\begin{aligned}
\lambda(p,c)=\frac{c}{p_{0}}\Bigg\{&\left[4\textrm{Re}(c)+(1-2p)^{2}\right]\frac{p_{0}}{\arctanh p_{0}}\\
&-4i\textrm{Im}(c)\frac{\arctanh p_{0}}{p_{0}}\Bigg\}
\label{lambda}
\end{aligned}
\end{equation}
If one notices that
$$
(f_{1}+f{_2})\rho_{t}-f_{1}\sigma_{1}^{z}\rho_{t}\sigma_{1}^{z}-f_{2}\sigma_{2}^{z}\rho_{t}\sigma_{2}^{z} = 
(f_{1}+f_{2})\left(
\begin{array}{cc}
0  &  4c  \\
4c^{*}  &  0 
\end{array}
\right)
$$
and renames $\gamma_{L}=\frac{4}{\beta}$ and $\gamma_{NL}=4\tau$, $|\downarrow,\uparrow\rangle\langle\uparrow,\downarrow|=S_{-}$ and $S_{+}=|\uparrow,\downarrow\rangle\langle\downarrow,\uparrow|$, Eq.~\eqref{generic2level} can finally be recast in the following very simple form
\begin{equation}
\dot{\rho_{t}}=i[\rho,H_{2}]-(f_{1}+f_{2})\left[\Gamma(p,c) S_{-}+\Gamma^{*}(p,c) S_{+}\right],
\label{generic2levelfinal}
\end{equation}
where 
$$
\Gamma(p,c)=4\gamma_{L}c+\gamma_{NL}\lambda(p,c),
$$
contains all the dissipation and decoherence. Thus, in the one-excitation subspace $S_{z}=0$, the original master equation \eqref{generic2} can be mapped onto a simpler master equation for an effective two-level system where the effect of the heat bath manifests itself as incoherent transitions deriving from both linear and nonlinear contributions. This makes an analytical time-dependent solution of Eq.~\eqref{generic2levelfinal} out of reach. However, as stated above, we can analytically solve $\dot{\rho_{t}}=0$. Assuming $\rho_{ss}$ of the same form as in Eq.~\eqref{rho0} one finds
\begin{equation}
\begin{cases}
c_{2}^{\textrm{ss}} = 0 \\
\textrm{Re}\Gamma(p^{\textrm{ss}},c_{1}^{\textrm{ss}},c_{2}^{\textrm{ss}})=0\\
(2p-1) - (f_{1}+f_{2})\textrm{Im}\Gamma(p^{\textrm{ss}},c_{1}^{\textrm{ss}},c_{2}^{\textrm{ss}})=0
\end{cases}
\label{equilibriumEq}
\end{equation}
where $c_{1}=\textrm{Re}(c), c_{2}=\textrm{Im}(c)$. By setting $c_{2}=0$ one finds that for any physical state $\textrm{Im}\Gamma(p^{\textrm{ss}},c_{1}^{\textrm{ss}},0)=0$ which implies $p^{\textrm{ss}}=1/2$. Finally, using the second equation, that is $\textrm{Re}\Gamma(1/2,c_{1}^{\textrm{ss}},0)=0$, one finds 
\begin{equation}
c_{1}^{\textrm{ss}}=-\frac{1}{2}\tanh\frac{\gamma_{NL}}{\gamma_{L}}.
\label{2spin_ss}
\end{equation} 
If one compute the subspace-restricted Gibbs state $\rho_{\textrm{G}}=e^{-\beta H}/\mathcal{Z}$, one will find that this exactly coincides with $\rho_{\textrm{ss}}$. Interestingly, the total steady-state energy is directly connected to the the real part of the coherence as follows
\begin{equation}
c_{1}^{\textrm{ss}}=\frac{\langle H \rangle_{\rho_\textrm{ss}}+1}{2}.
\label{c1(E)}
\end{equation}

\section{Many-body thermalization}
\label{sec:solution}
In this section we are going to numerically solve the thermodynamic master equations \eqref{generic2} and \eqref{FinalGenericEq} for $N=2$ and $N=10, 20, 30, 40$ spins, respectively. From now on, $f_{j}=\gamma$, for all $\alpha$, and $\tau=1$. In all cases considered in this manuscript, the GENERIC master equation was solved exactly using an implicit Runge-Kutta method, leading to excellent results with reasonable computational resources. 

\begin{figure}[htbp]
\begin{center}
\includegraphics[width=0.4\textwidth]{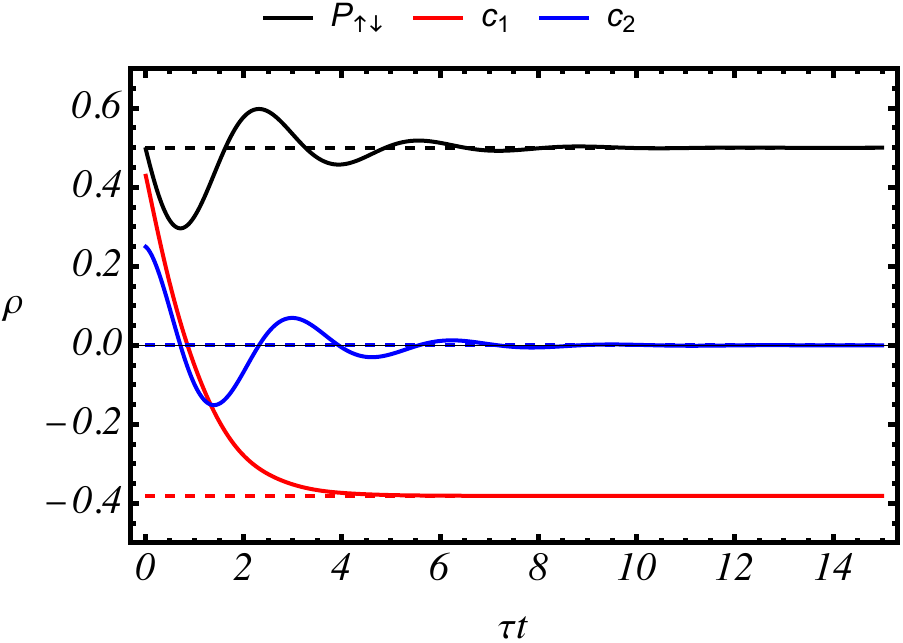}
\caption{$N=2$: Time-evolution of $P_{\uparrow,\downarrow}$ (black), $c_{1}$ (red) and $c_{2}$ (blue) obtained from the numerical solution of Eq.~\eqref{generic2} (continuous) in comparison to their steady-state values (dashed) as given by the effective two-level model. The initial state is $|\psi_{0}\rangle=(|\downarrow,\uparrow\rangle+e^{i\pi/6}|\uparrow,\downarrow\rangle)/2$, while the other parameters are $\gamma=0.05, \beta=1$.}
\label{2spinRho}
\end{center}
\end{figure}

For the $N=2$ case, we are going to use the analytically calculated steady-state solution of Eq.~\eqref{generic2levelfinal}, that is the correct Gibbs state, and compare it with the numerically obtained long-time solution of Eq.~\eqref{generic2}. In Fig.~\ref{2spinRho} the time evolution of the density matrix diagonal and off-diagonal elements (continuous) is shown as obtained by numerically solving Eq.~\eqref{generic2} for an initial pure state. The steady-state values calculated analytically are also shown (dashed). The agreement between numerical results and steady-state theoretical predictions is excellent and was verified for every initial state used.

For the case of $N$ spins, the complexity of Eq.~\eqref{FinalGenericEq} makes any analytical calculation out of reach. Thus, in this case, we proceed differently. We first calculate numerically the Gibbs state $\rho_{G}$ as in Eq.~\eqref{gibbs}. We then solve our thermodynamic master equation \eqref{FinalGenericEq} and compare the long-time solution to $\rho_{G}$. In order to gain a clearer understanding of the different time-scales that might be involved in the thermalisation process we look at two different quantities. 

\begin{figure}
     \centering
     \begin{subfigure}[b]{0.45\textwidth}
         \centering
         \includegraphics[width=\textwidth]{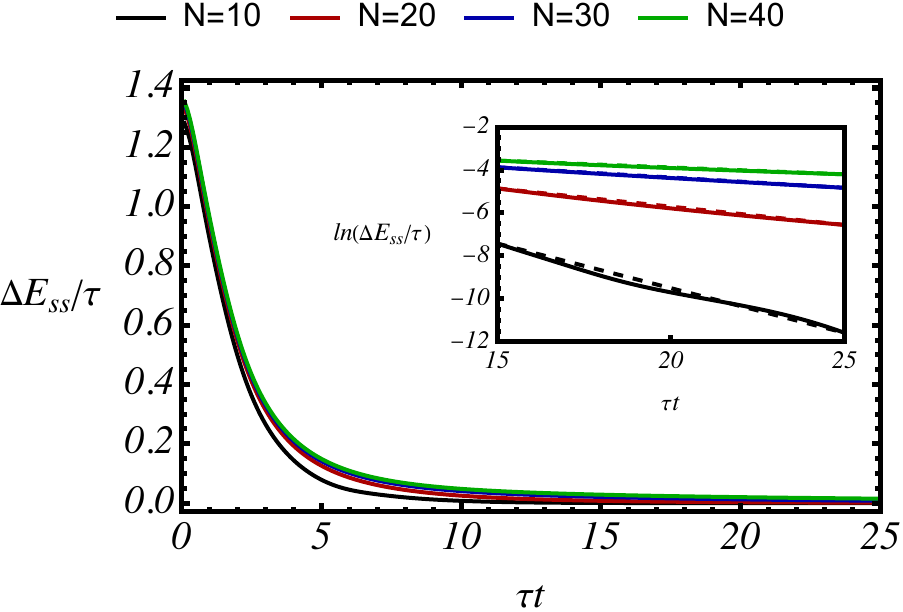}
         \label{energy_eq}
     \end{subfigure}
  \begin{subfigure}[b]{0.45\textwidth}
         \centering
         \includegraphics[width=\textwidth]{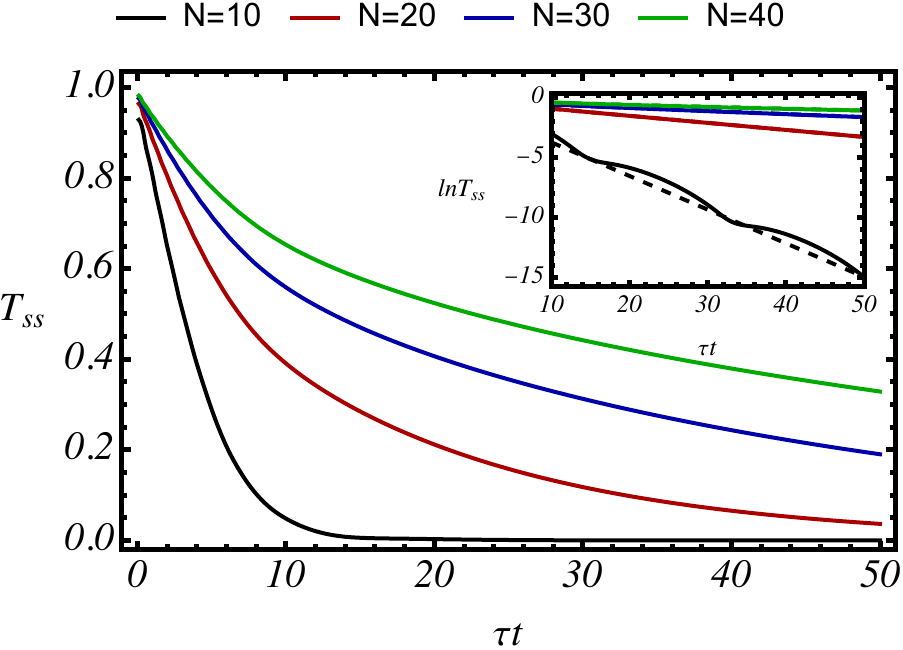}
         \label{trace_eq}
     \end{subfigure}
     	   \caption{Top: Time-evolution of $\Delta E_{ss}/\tau$ for different chain of length $N=10$ (black), $N=20$ (darker red), $N=30$ (darker blue), $N=40$ (darker green), as obtained from the numerical solution of Eq.~\eqref{FinalGenericEq}. The initial state is $|\psi_{0}\rangle=|\uparrow_{1}\rangle\otimes|\ \Downarrow\rangle_{N-1}$, while the other parameters are $\gamma=0.05, \beta=1$. Bottom: Time-evolution of $T_{ss}$ for the same parameters.}
            \label{mb_eq}
\end{figure}

First, we study energy equilibration via the following energy difference
\begin{equation}
\Delta E_{ss} = \textrm{Tr}[H(\rho_{t}-\rho_{G})].
\label{deltaE}
\end{equation}
The initial state of the chain is $|\psi_{0}\rangle=|\uparrow_{1}\rangle\otimes|\downarrow_{2},\downarrow_{3},\dots,\downarrow_{N}\rangle$, that is, a spin-flip, or excitation, localised at one end of the chain. We remind that the $S_{z}$ is conserved and thus, in solving Eq.~\eqref{FinalGenericEq}, we can restrict our attention to the subspace spanned by similar single spin-flip states, that is, states of the form 
$ |n\rangle \equiv \sigma_{n}^{+}|\downarrow,\downarrow\dots,\downarrow\rangle=\sigma_{n}^{+}|\Downarrow\rangle_{N}$, where we have introduced the shorthand notation $|\downarrow,\downarrow\dots,\downarrow\rangle=|\Downarrow\rangle_{N}$.
The top panel of Fig.~\ref{mb_eq} shows the energy equilibration for different chain sizes, that is, $N=10, 20, 30, 40$. As expected, all the four lines vanish in the long-time limit with all the lines becoming more closely packed when increasing $N$. In the inset we focus specifically on the long-time behaviour and display $\ln\Delta E_{ss}/\tau$. At larger $N$ the time-evolution appears approximately linear, which is consistent with a long-time limit exponential decay $\Delta E_{ss}\propto e^{-\Gamma_{E} t}$. 

\begin{figure}[htbp]
\begin{center}
\includegraphics[width=0.42\textwidth]{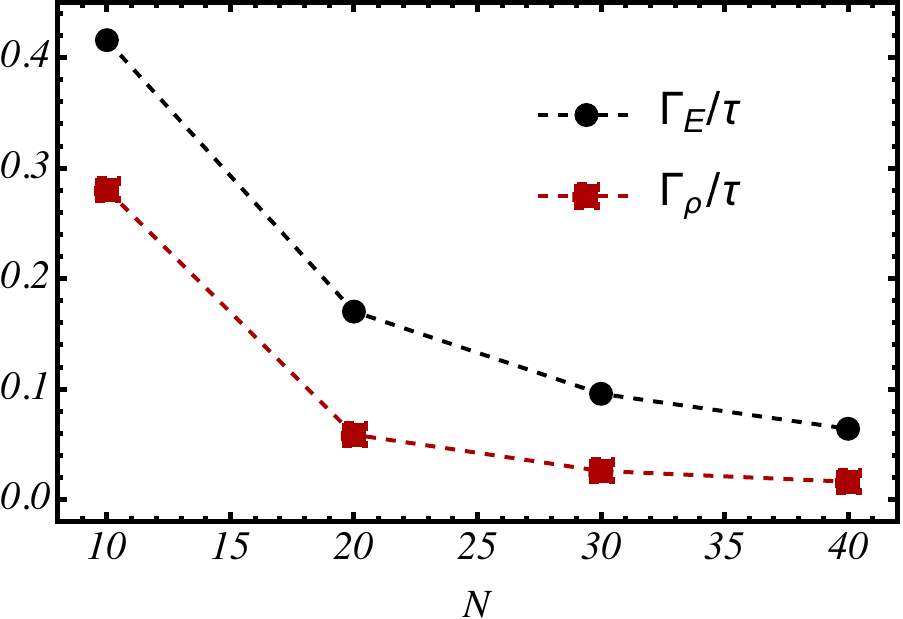}
\caption{Rescaled energy and state thermalisation rates $\Gamma_{E}$ and $\Gamma_{\rho}$ for different chain lengths $N$, as obtained from linear extrapolation of $\ln\Delta E_{ss}$ and $\ln T_{ss}$.}
\label{gamma}
\end{center}
\end{figure}

To investigate thermalisation at the quantum state level we use the trace distance $T_{ss}(t)$ between $\rho_{G}$ and $\rho_{t}$, which is defined as
\begin{equation}
T_{ss}\equiv\frac{1}{2}\textrm{Tr}\left[\sqrt{(\rho_{t}-\rho_{G})^{2}}\right]=\frac{1}{2}\sum_{j}|\lambda_{j}(t)|,
\label{traced}
\end{equation}
where $\lambda_{j}(t)$ are the time-dependent eigenvalues of $\rho_{t}-\rho_{G}$. The trace distance $T(\rho,\nu)$ is bounded, that is $0\le T\le 1$, symmetric and $T(\rho,\nu)=0$ iff $\rho=\nu$. As such, it is a very good measure to distinguish quantum states. The bottom panel of Fig.\ref{mb_eq} displays the process of thermalisation as measured by the trace distance $T_{ss}$ again for $N = 10, 20, 30, 40$ spins. All the curves decay in the long-term limit, however, at a smaller rate. Similarly to the case of energy, in the inset of we show $\ln T_{ss}$. Again, at larger $N$ and in the long-time limit, we see an almost perfectly linear behaviour, which translates to a nearly exact exponential decay $T_{ss}\propto e^{-\Gamma_{\rho} t}$ with a decay rate $\Gamma_{\rho}$ smaller than $\Gamma_{E}$. We notice that, while for $\ln \Delta E_{ss}$ a tiny deviation from perfect linear behaviour can be still observed, the same does not apply to $\ln T_{ss}$, which is virtually indistinguishable from its linear interpolation at larger $N$. 
Finally, in Fig.~\ref{gamma} we show both $\Gamma_{E}$ and $\Gamma_{\rho}$ versus $N$, as obtained from linear interpolation. Both quantities are monotonically decreasing functions of $N$ as one would expect when increasing the size of a many-body system. Although for the $N$ range used here $\Gamma_{\rho}< \Gamma_{E}$, the gaps between these two quantities also appear to decrease as with increasing $N$ and to indicate that $\Gamma_{\rho}\to\Gamma_{E}$ in the limit $N\to\infty$. 

We conclude this section by stressing again that all these results were found by numerically solving the full, nonlinear GENERIC equation \eqref{FinalGenericEq} without extra approximations and, as such, they are numerically exact solutions proving thermalisation within the $S_{z}=2-N$ subspace of the total Hilbert space. Moreover, if the nonlinearity was not included the resulting master equation would be a Lindblad dephasing-type master equation, whose stationary state is maximally mixed, that is $\rho_{M}=\mathcal{I}/2^{N}$, corresponding to the infinite temperature thermal state regardless of the system's Hamiltonian.

\section{Out-of-equilibrium many-body properties}
\label{sec:ooe}
In this section we focus on an out-of-equilibrium scenario. The Hamiltonian we consider reads
\begin{equation}
H = \tau\sum_{j = 1}^{N} \left(\sigma_{j}^{+}\sigma_{j+1}^{-}+\sigma_{j}^{-}\sigma_{j+1}^{+}\right)+\Delta \sigma_{j}^{z}\sigma_{j+1}^{z} +\frac{\Omega}{2}\sigma_{j}^{z},
\label{simpler_H_neq}
\end{equation}
where we added a single-spin local term. Here, the extremities of the spin chain exchange spin excitations locally with two separated environments, say $L$ and $R$, whilst the bulk of the chain is isolated. This type of model has been extensively studied in literature  in connection with nonequilibrium steady states and transport properties and an abundance of both analytical and numerical results are available \cite{Michel:2003vv,PhysRevE.86.061118,PhysRevB.80.035110,PhysRevB.87.235130,Mendoza_Arenas_2013,PhysRevLett.95.180602,PhysRevE.76.031115,Prosen_2008,Prosen_2009,_nidari__2010,PhysRevLett.107.137201,PhysRevLett.124.160403}. Generally speaking, it is commonly accepted that a transition from ballistic ($\Delta<1$) to diffusive transport ($\Delta>1$) occurs at $\Delta=1$ both in the linear perturbation \cite{PhysRevB.55.11029,Kl_mper_2002} and in the open quantum systems framework \cite{PhysRevLett.106.217206,PhysRevLett.107.137201}.

Here, we model the two environments as heat baths at temperatures  $T_L$ and $T_R$ respectively and model the system-bath coupling with $\sigma_{\pm}$. Since $S_{z}$ is no longer conserved we are now forced to take into account the full Hilbert space in solving the GENERIC master equation and this, combined with the non-linear nature of our thermodynamics approach, drastically limits the system sizes that are accessible to an exact numerical solution. Thus, for this work, we restrict our attention to small chains with $N=3,4,5$. Fortunately, if the deviation from equilibrium is not too drastic one can replace the nonlinear terms $\rho_{t}^{u(1-u)}$ with $\rho_{G}^{u(1-u)}$ \cite{HansZeroTemp,PhilApp}. The resulting equation will be thus linear in $\rho_{t}$, leading to a considerable computational speed-up. In the context of interacting many-body systems this paves the way to the development of a stochastic unravelling that would allow for a system-size scale-up, an approach that we are currently investigating. 

Going back to the full nonlinear equation, starting from Eq.~\eqref{generalME} and similarly to the derivation in Sec.\ref{sec:model}, one obtains the following master equation 
\begin{equation}
\begin{aligned}
\dot{\rho_{t}} =  i\left[\rho_{t},H\right] + \sum_{\alpha=L,R}\sum_{\mu=\pm}\mathcal{L}_{j}^{\mu}[\rho_{t}]+\mathcal{N}_{j}^{\mu}[\rho_{t}],
\end{aligned}
\label{non-eq-Generic}
\end{equation}
where we have renamed the boundary spin indexes as $1=L$ and $N=R$, the linear part of the dissipator reads
\begin{equation}
\mathcal{L}_{j}^{\mu}[\rho_{t}]=\gamma_{\mu}\left(\sigma_{j}^{\mu}\rho_{t}\sigma_{j}^{\mu\dagger}-\frac{\{\sigma_{j}^{\mu\dagger}\sigma_{j}^{\mu},\rho_{t}\}}{2}\right),
\label{lind}
\end{equation}
which is a standard local Lindblad dissipator with rates satisfying detailed balance {\itshape i.e.} $\gamma_{+}=e^{-\beta_{j}\Omega}\gamma_{-}=e^{-\beta_{j}\Omega}\gamma$, while the nonlinear contribution is modelled as
\begin{equation}
\mathcal{N}_{j}^{\mu}[\rho_{t}]=\gamma\beta_{j}\int du \;e^{-\beta\Omega u}[\sigma_{j}^{\mu},\rho_{t}^{\eta_{\mu}(u)}[\sigma_{j}^{\mu\dagger},V]\rho_{t}^{1-\eta_{\mu}(u)}]
\label{non_lind}
\end{equation}
where $\eta_{-}(u)=1-u=1-\eta_{+}(u)$. The exponentials $e^{-\beta\Omega u}$ are introduced to guarantee detailed balance. Similarly to Eq.~\eqref{nonLinear} the above nonlinear contribution can be numerically calculated by expanding the integrand over the density matrix eigenstates at all times. 

In the long-time limit we should expect the system to reach a nonequilibrium steady state, say $\rho_{ness}$ with certain energy and energy current distributions along the chain. In order to study the energy properties of $\rho_{ness}$ we numerically solve the above equation assuming the XXZ ground state $|GS\rangle$ as an initial state. The extremities of the chain are then connected to the two different heat baths and let interact, leading to a redistribution of the initial energy which will depend on the interplay between the Hamiltonian $H_{XXZ}$ and the effect of both the baths as well as the temperature gradient.
We use the following definition of local energy density \cite{Mendoza_Arenas_2013} 
\begin{equation}
\epsilon_{j}=\langle\tau(\sigma_{j}^{+}\sigma_{j+1}^{-}+\textrm{h.c})+\Delta\sigma_{j}^{z}\sigma_{j+1}^{z} + \frac{\Omega}{2}\sigma_{j}^{z}\rangle=\langle h_{j}\rangle
\label{local_energy}
\end{equation}
The energy current can be derived easily using a continuity equation for the energy profile \eqref{local_energy} \cite{Mendoza_Arenas_2013}. For $2\le\alpha\le N-1$ this reads
\begin{equation}
\dot{\epsilon}_{j} = i\langle[H,h_{j}]\rangle=-\nabla\cdot \langle J_{j}\rangle=\langle J_{j}\rangle-\langle J_{j+1}\rangle,
\label{cont_eq_energy}
\end{equation}
which, $h_{j} = [h_{\alpha-1},h_{j}]+[h_{j+1},h_{j}]$, allows one to define
\begin{equation}
\langle J_{j}\rangle = J^{\textrm{BULK}} =i\langle[h_{\alpha-1},h_{j}]\rangle.
\label{j_along_chain}
\end{equation}
where we have tacitly assumed a uniform energy current, which will turn out to be the case.
\begin{figure}[htbp]
\begin{center}
\includegraphics[width=0.45\textwidth]{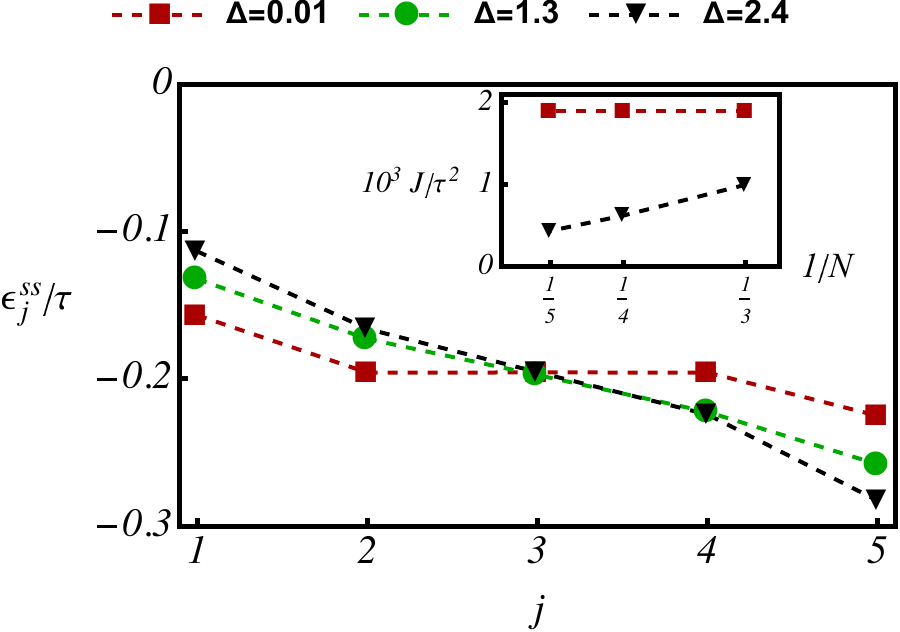}
\caption{Local steady-state energy density $\epsilon_{j}^{ss}$ for $N=5$ at different couplings $\Delta = 0.01, 1.2, 2.4$  or $\beta_{L}=0.41$ and $\beta_{R}=1.39$. Inset: steady-state current as a function of $N=3,4,5$ at $\Delta = 0.01$ and $\Delta = 2.4$.}
\label{eps_i}
\end{center}
\end{figure}
\begin{figure}[htbp]
\begin{center}
\includegraphics[width=0.46\textwidth]{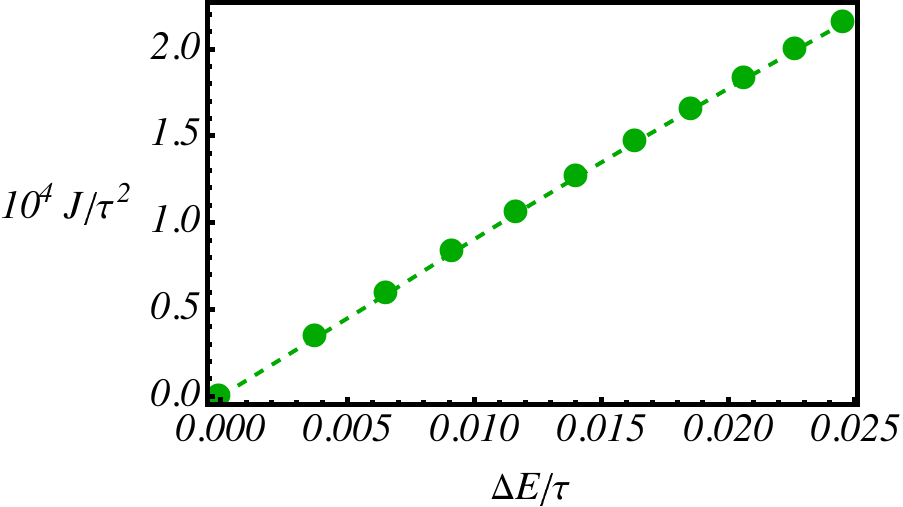}
\caption{Steady-state energy current $J^{\textrm{BULK}}$ as a function of the energy gradient $\Delta E$ across the chain for $N=4$ at $\Delta=1.2$.}
\label{jnablaE}
\end{center}
\end{figure}

In the case of ballistic transport at $\Delta<1$, one should expect a nearly flat local energy distribution in the bulk of the chain while in the diffusive regime at $\Delta>1$ a constant energy gradient should emerge. In the top panel of Fig.~\ref{eps_i} we display $\epsilon_{j}$ for $N=5$ at three different values of the anisotropy parameter $\Delta = 0.01,1.2,2.4$ corresponding to low, strong and ultra-strong coupling. A transition from a flat to a linearly decreasing energy density can be clearly observed. This result is further corroborated by looking at the bulk steady-state current as a function of the number of spins, which is shown in the inset panel for $N=3,4,5$. For $\Delta<1$ the steady-state current is independent of the system size, while in the diffusive regime it decays as $1/N$. 
This is fully consistent with Fourier's law of heat transport which for a finite system reads
\begin{equation}
J^{\textrm{BULK}} = \kappa \frac{\Delta E}{N},
\label{fourier}
\end{equation}
with $\kappa$ being the conductivity and $\Delta E$ the energy difference across the chain. We conclude this section by showing the dependence of $J^{\textrm{BULK}}$ on this energy difference for $N=4$ strongly interacting spins in Fig.~\ref{jnablaE}, where a linear character can be seen clearly, in agreement with Eq.~\eqref{fourier}. 

\section{Conclusions and open perspectives.}
\label{sec:conclusions}
In this manuscript we used a nonlinear quantum thermodynamics master equation, named GENERIC, to study equilibrium and out-of-equilibrium dynamics in a spin chain. We were able to obtain numerically exact solutions for i) long chains in the equilibrium scenario and ii) shorter chains in the nonequilibrium one. In the case of equilibration in a heat bath we found that the density matrix of the many-body system approaches to correct interacting many-body Gibbs state in the long-time limit and estimated two different thermalization rates describing energy and state dynamics, respectively. In the nonequilibrium scenario, we demonstrated ballistic to diffusive transport and recovered the Fourier's law of heat transport in agreement with previous literature.

Certainly, a majorly important aspect of GENERIC is that, under fairly general assumptions (\textit{e.g.}, KMS condition), it allows one to study both equilibrium and nonequilibrium scenarios without having to invoke all those approximations that are usually required using a Lindblad/Redfield approach. On the other hand, GENERIC does require extra modelling that should be guided by thermodynamic principles and that, in some cases, might no be obvious a priori. While we acknowledge that traditional approaches based on Redfield/Lindblad are indeed very powerful tools, we believe that the findings presented in this manuscript certainly pave the way for further investigations and/or applications of this alternative approach as well, {\textit e.g.}, in quantum thermodynamics and quantum thermometry\\
\section{Acknowledgments}
M.B. would to thank David Taj for the interesting discussion clarifying the connections between GENERIC and Lindblad master equations and Rea Dalipi for useful discussions on the Bethe Ansatz in 1D systems.

\bibliography{biblio.bib}

\end{document}